\documentstyle[amssymb,prd,aps,twocolumn]{revtex} 

\newcommand {\beq} {\begin{equation}}
\newcommand {\eeq} {\end{equation}}
\newcommand {\beqa}{\begin{eqnarray}}
\newcommand {\eeqa}{\end{eqnarray}}
\newcommand {\del} {\partial}
\newcommand {\rmd} {{\rm d}}

\newcommand {\CS}     {Chern--Simons}
\newcommand {\bc}     {boundary condition}
\newcommand {\bcs}    {boundary conditions}
\newcommand {\pbc}    {periodic boundary condition}

\newcommand {\rhs}    {right-hand side}
\newcommand {\cdet}   {chiral determinant}
\newcommand {\cdets}  {chiral determinants}

\newcommand {\cgths}  {chiral gauge theories}

\begin{document}
\draft

\title{CPT anomaly in two-dimensional chiral $U(1)$ gauge theories}

\author{F.R. Klinkhamer$^{a}$\cite{EmailFRK}
        and J. Nishimura$^{b}$\cite{EmailJN}}
\address{$^{a}$ Institut f\"ur Theoretische Physik, Universit\"at Karlsruhe,
D--76128 Karlsruhe, Germany}
\address{$^{b}$ The Niels Bohr Institute,
Blegdamsvej 17, DK-2100 Copenhagen \O, Denmark}

\date{
      hep-th/0006154; Phys. Rev. D 63, 097701 (2001)}

\twocolumn[\hsize\textwidth\columnwidth\hsize\csname@twocolumnfalse\endcsname

\maketitle
\begin{abstract}
The CPT anomaly, which was first seen
in  perturbation theory for certain four-di\-men\-si\-o\-nal chiral gauge theories,
is also present in the exact result for a class of two-dimensional chiral $U(1)$
gauge theories on the torus.
Specifically, the chiral determinant for periodic fermion
fields changes sign under a CPT transformation of the background gauge field.
There is, in fact,  an anomaly of Lorentz invariance, which
allows for the CPT theorem to be circumvented.
\end{abstract}

]   

\section{Introduction}
Recently, a CPT anomaly has been found in certain four-dimensional
\cgths, with the topology and spin structure of the spacetime
manifold playing a crucial role \cite{K00}.
The well-known CPT theorem \cite{L57} is circumvented by the breakdown of
Lorentz invariance at the quantum level \cite{K00,K98}.
The calculation of Ref. \cite{K00} was done perturbatively and
more or less the same type of anomaly was expected to occur
in appropriate higher- and lower-dimensional \cgths.
Here, we consider the
two-dimensional chiral $U(1)$ gauge theory over the torus, for which the
\cdet~is known \emph{exactly} \cite{AMV86,KNN97,IN99}.
The aim of this paper is to determine
whether or not the exact result contains the CPT anomaly
and perhaps to learn more about the anomaly itself \cite{note2D}.

\section{Chiral determinant}
We consider in this Brief Report two-dimensional Euclidean chiral $U(1)$ gauge theory,
defined over the torus $T^2$.
For simplicity, we take a particular torus (modulus $\tau = i$),
with Cartesian coordinates $x^\mu$ $\in$ $[0,L]$,
$\mu=1$, 2,  and Euclidean metric $g_{\mu\nu}$ $=$ $\delta_{\mu\nu}$.
The theory has the fermionic action
\beq \label{eq:action}
S[A,\bar{\psi},\psi] =
                      - \int_{0}^{L} \rmd  x^1 \int_{0}^{L}\rmd  x^2 \;\bar{\psi }\,
                        \sigma^\mu \!\left(\del _\mu + i A_\mu  \right)  \psi \;,
\eeq
with $\sigma^1 = 1$ and $\sigma^2 = i$. The boundary conditions for the
real gauge potential $A(x)$ $\equiv$ $A_\mu(x)\,\rmd  x^\mu$
and the 1-component Weyl field $\psi(x)$ are \emph{both} taken to be
periodic:
\begin{eqnarray}  \label{eq:bcs}
           A     (x^1 + m L, x^2 + n L) & = & A     (x^1,x^2)\;,\nonumber\\
           \psi  (x^1 + m L, x^2 + n L) & = & \psi  (x^1,x^2)\;,
\end{eqnarray}
for  arbitrary integers $m$ and $n$.

The two-dimensional gauge potential in the trivial topological sector
can be decomposed as follows \cite{AMV86}:
\beq
A_\mu (x)= \epsilon_{\mu\nu}\, g^{\nu\rho}\,\del _\rho \phi (x)  +
           2\pi h_\mu/L + \, \del_\mu \chi (x)  \;,
\label{eq:Adecomposed}
\eeq
with $\phi(x)$ and $\chi(x)$  real periodic functions
and $h_\mu$ real constants (the harmonic pieces of the
gauge potential). Here, $\chi(x)$ corresponds to the gauge degree of
freedom. Furthermore, the gauge potential
$A_\mu(x)$ is taken to be smooth, i.e. without delta-function singularities.

The \cdet~(the exponential of minus the Euclidean effective action) is then
given by the following functional integral:
\beqa
D^{\rm PP}[A]   &\equiv& \exp \left( - \Gamma^{\rm PP}[A] \right) \nonumber \\
                &  =   & \int_{\rm PP} {\cal D} \psi \, {\cal D} \bar{\psi}~
                         \exp \left(- S[A, \bar{\psi},\psi ]\right)  \;,
\label{eq:pathintegral}
\eeqa
where PP indicates the doubly-\pbc~(\ref{eq:bcs}) on the fermion field.
This \cdet~has been calculated using various regularization methods.
See Refs. \cite{AMV86,KNN97,IN99} and references therein.
Reference \cite{IN99}, in particular, introduces a local
counterterm to restore translation invariance and obtains the
following result \cite{noteDPP}:
\beqa \label{eq:DPP}
D^{\rm PP}[A] &=&
\hat{\vartheta} (h_1 + {\textstyle\frac{1}{2}},
                                   h_2 + {\textstyle\frac{1}{2}} )\,
  \exp \left( \,\frac{i\pi}{2} (h _ 1 - h _2)  \right) \nonumber \\
&~&
 \times  \exp \left( \frac{1}{4 \pi}
    \int \rmd ^2 x \left( \phi \,\del ^2 \phi + i \phi \,\del ^2 \chi  \right)
        \right),
\eeqa
with, for real variables $k_1$ and $k_2$, the definition \cite{KNN97,IN99}
\beqa
\hat{\vartheta} (k_1,k_2) &\equiv&
                          \exp\left[ - \pi (k_2)^2 + i \pi k_1 k_2\right]
\nonumber \\
                          && \times \, \vartheta (k_1+i k_2;i)/ \eta (i)  \; ,
\eeqa
in terms of the Riemann theta function and Dedekind eta function
\beqa
\vartheta (z;\tau) &\equiv& \sum_{n=-\infty}^{\infty}
                   \exp\left(\pi  i n^2  \tau + 2 \pi i n z \right),\nonumber\\
\eta (\tau )       &\equiv& \exp\left(\pi i \tau/12\right) \prod _{m=1} ^{\infty}
                   \left[1 - \exp\left(2\pi i m \tau \right)\right]\;.
\eeqa

The result (\ref{eq:DPP}) holds for the \cdet~of a single positive
chirality (right-moving) Weyl fermion of
unit charge; cf. Eq. (\ref{eq:action}). If the charge is $q_{R1}$ instead,
then the  variables
$h_\mu$, $\phi(x)$, and $\chi(x)$ in Eq. (\ref{eq:DPP}) each need to be multiplied by
a factor $q_{R1}$. For a negative chirality
(left-moving) Weyl fermion of charge $q_{L1}$, one
also has to take the complex conjugate of the whole expression (\ref{eq:DPP}). For the
345-model (three chiral fermions with charges $q_{R1}=3$, $q_{R2}=4$, and $q_{L3}=5$),
one obtains the following \cdet~\cite{IN99}:
\beq \label{eq:D345}
D^{\rm PP}_{345}[A] = D^{\rm PP}[3A] \, D^{\rm PP}[4A]\,
                      \left(D^{\rm PP}[5A] \right)^* \;.
\eeq
The \cdet~(\ref{eq:D345}) of the 345-model
is gauge invariant. Indeed, it is straightforward to verify
both the $\chi$ independence
and the invariance under large gauge transformations
$h_\mu$ $\rightarrow$ $h_\mu+n_\mu$ for arbitrary integers
$n_\mu$ \cite{notePhase}. We will first focus on this particular chiral model.
Other chiral models will be discussed later.

\section{CPT noninvariance}
The question, now, is how the gauge-invariant \cdet~(\ref{eq:D345})
of the 345-model behaves under a CPT transformation of the background gauge
field:
\begin{equation}\label{eq:ACPT}
A_\mu(x) \, \rightarrow \,A^{\rm CPT}_\mu(x) \equiv - A_\mu(-x)\; .
\end{equation}
Using the elementary properties of the theta function \cite{noteTheta},
one finds
\beq \label{eq:345CPT}
D^{\rm PP}_{345}[A^{\rm CPT}] = - D^{\rm PP}_{345}[A]\;,
\eeq
with each of the three chiral fermions contributing a multiplicative
factor $-1$ on the \rhs. Hence, the effective action of the chiral
$U(1)$ gauge theory with PP spin structure over the torus
\emph{changes} under a CPT transformation (\ref{eq:ACPT}) of the background gauge
field, provided the total number ($N_F$) of charged chiral fermions of the
theory is \emph{odd} (e.g. $N_F$ $=$ $3$ for the 345-model).
The result (\ref{eq:345CPT}) thus
provides conclusive evidence for a CPT anomaly
of the chiral model considered.

The asymmetry (\ref{eq:345CPT}) implies the vanishing of the
\cdet~(\ref{eq:D345}) for $A_\mu(x)$ $=$ $0$.
For gauge fields (\ref{eq:Adecomposed}) with
$\phi(x)$ $=$ $\chi(x)$ $=$ $0$
and infinitesimal harmonic pieces $h_\mu$, one has, in fact,
\beq \label{eq:D345linear}
D^{\rm PP}_{345}[h_1,h_2] =
c\,(h_1 + i h_2)\,(h_1^2 + h_2^2) + {\rm O}(h^5) \;,
\eeq
with a nonvanishing complex constant $c$. This result follows from the
observation that
the analytic function $\vartheta(z;i)$ appearing in Eq. (\ref{eq:DPP})
has a simple zero at $z$ $=$ $(1+i)/2$.
More directly, the holomorphic factor $(h_1 + i h_2)$ in Eq. (\ref{eq:D345linear})
corresponds to one of the eigenvalues  of the Weyl operator
$\sigma^\mu \left( \del_\mu + i A_\mu \right)$  with  doubly-periodic \bc~and
constant gauge potential, as do the holomorphic and antiholomorphic factors
contained in $(h_1^2 + h_2^2)$.
Equation (\ref{eq:D345linear}) agrees, of course, with the general result
(\ref{eq:345CPT}) on CPT violation.
But the real importance of Eq. (\ref{eq:D345linear}) is
that, for this special case, the \emph{origin}
of the two-dimensional
CPT anomaly can be identified explicitly, namely one particular
eigenvalue of the Weyl operator. (See \cite{noteEigenval} for further details.)

The \cdet~\cite{IN99} of the 345-model over the torus is CPT
invariant for the other spin structures AA,
PA, and AP, where (A)P stands for (anti-)periodic
\bcs~on the fermion fields (the three fermion species being treated equally).
This appears to be related to the observation that the CPT anomaly is
not expected for the AA spin structure
\cite{K00,K98} and the fact that the \cdets~\cite{IN99} for the
AA, PA, and AP spin structures transform into each
other under modular transformations
(global diffeomorphisms; cf. Ref. \cite{AMV86}),
whereas the \cdet~of the PP spin structure is invariant up to a phase.
It is important to realize that this extra
requirement of modular invariance for the AA, PA,
and AP spin structures restricts the type of
theories considered and also possible
regularization methods \cite{notePV}.
For the general question of
how to sum over the different spin structures, see,
for example, the discussion in Refs.
\cite{AI79,GSW87}. In our case, the two-dimensional
CPT anomaly would be present as long as the PP spin
structure appears in the sum.

\section{Lorentz noninvariance}
Given that CPT invariance no longer holds for the 345-model with
doubly-periodic spin structure over the torus, $SO(1,1)$
Lorentz invariance, or rather $SO(2)$ invariance for the Euclidean
theory, is expected to be broken as well \cite{K00,K98}.
Concretely, this can be tested by comparing the (translation-invariant)
\cdet~(\ref{eq:D345}) for two different, \emph{localized} gauge fields
which are related by a Lorentz transformation \cite{CK97}.

Consider, for example, a gauge potential $\tilde{A}_\mu(x)$
which, up to periodicity, is allowed to be
nonzero only for $|x^\mu - L/2|$ $<$ $\ell$,
with a fixed length $\ell$ $<<$ $L/2$,
and which has infinitesimal, but nonvanishing,
harmonic pieces $\tilde{h}_\mu$ $\equiv$
$(2\pi L)^{-1} \int \rmd^2 x \,\tilde{A}_\mu(x)$.
In other words, the gauge potential $\tilde{A}_\mu(x)$ has local support
(set by $\ell$) and produces small, but nonzero, averages $\tilde{h}_\mu$
(typically of order $\ell /L$).
According to Eq. (\ref{eq:D345linear}), the \cdet~for this gauge field is
then proportional to $(\tilde{h}_1 + i \tilde{h}_2)$
$=$ $\sigma^\mu \,\tilde{h}_\mu$. Similarly, the \cdet~for
the $SO(2)$  Lorentz transformed  (``boosted'') gauge potential,
\beq \label{eq:Arotation}
\left( \begin{array}{c}
       \tilde{A}_1^\prime (x)\\
       \tilde{A}_2^\prime (x) \end{array}\right) = \Lambda \cdot
\left( \begin{array}{c} \tilde{A}_1 (\Lambda\, x)\\ \tilde{A}_2 (\Lambda\, x)
       \end{array}\right),
\;\;    \Lambda  \equiv \left( \begin{array}{cc}
            \cos\alpha  & -\sin\alpha \\
            \sin\alpha &  \cos\alpha   \end{array} \right) ,
\eeq
is proportional to $\sigma^\mu \,\tilde{h}^\prime_\mu$.
But these two particular factors differ by a phase factor $\exp (i
\alpha)$, as can be readily verified.
All other factors of the two \cdets~being equal, this then implies
\beq \label{eq:345Lorentz}
     D^{\rm PP}_{345}[\tilde{A}^\prime] =  \exp (i \alpha) \,
                                           D^{\rm PP}_{345}[\tilde{A}]\;.
\eeq
Note that Eq. (\ref{eq:345Lorentz}), for $\alpha$ $=$ $\pi$, agrees with the
previous result (\ref{eq:345CPT}).
Also note that the noninvariance of the factor $\sigma^\mu \,\tilde{h}_\mu$ in
the \cdet~directly carries over to the theory with Minkowskian
metric $g_{\mu\nu}$ $=$ $\mathrm{diag}$ $(+1,-1)$.
In short, the Lorentz invariance of the \cdet~(\ref{eq:D345}) for
the localized gauge field $\tilde{A}_\mu(x)$
is broken through its $\tilde{h}_\mu$ dependence.
(The term $\int \rmd ^2 x \,\tilde{\phi} \,\del ^2 \tilde{\phi}$
from Eq. (\ref{eq:DPP}) is, of course, Lorentz invariant.)

As far as the gauge potential is concerned, this localized configuration
$\tilde{A}_\mu(x)$
could also have been embedded in the Euclidean plane ${\mathbb{R}}^2$.
The Lorentz noninvariance of the effective gauge field action comes
from the chiral fermions which are sensitive to the topology of the torus $T^2$.
More physically, the periodic \bcs~\emph{predispose} the chiral fermions
of the 345-model to select specific $\tilde{h}_\mu$--dependent
terms from the local interaction with the gauge field.
These special terms in the effective action then make the local dynamics of the
(classical) gauge field $\tilde{A}_\mu(x)$ Lorentz noninvariant.

\section{Other chiral models}
Up until now, we have focused on the 345-model,
which has an odd number of charged chiral fermions ($N_F$ $=$ $3$).
A chiral model with even $N_F$ does not have the CPT anomaly
discussed above, but can still be Lorentz noninvariant. An example
for $N_F$ $=$ $10$ would be the $1^9 3$-model, which has
ten chiral fermions with charges $q_{Ri}=1$, for $i=1$, $\ldots$ , $9$, and
$q_{L10}=3$. For this model, the chiral determinant (\ref{eq:D345linear})
becomes
\beq \label{eq:D1^93linear}
D^{\rm PP}_{1^9 3}[h_1,h_2] =
c^\prime \,(h_1 + i h_2)^8\,(h_1^2 + h_2^2) + {\rm O}(h^{12}) \;,
\eeq
which is invariant under the CPT transformation (\ref{eq:ACPT}),
but changes under the $SO(2)$ Lorentz transformation
(\ref{eq:Arotation}) by a phase factor
$\exp (i 8\alpha)$. On the other hand, a chiral model with even $N_F$
can also be Lorentz invariant, in the sense discussed above.
An example would be the chiral model with
$N_F=6$ chiral fermions of charges
$\{q_{R}\}$ $=$ $\{3,4,13\}$ and $\{q_{L}\}$ $=$ $\{5,5,12\}$, for which
the \cdet~is $c^{\prime\prime}\, (h_1^2 + h_2^2)^3 $ to lowest order.
(Vectorlike models, which have $\{q_{R}\}$ $=$ $\{q_{L}\}$,
are always Lorentz invariant.) Clearly, a deeper understanding
of what distinguishes these gauge-invariant chiral models remains to be
desired.

\section{Discussion}
For the two-dimensional
chiral $U(1)$ gauge theory with an odd number $N_F$
of charged chiral fermions defined over the torus, we have thus seen
that the CPT noninvariance of the effective gauge field action
$\Gamma^{\rm PP}[A]$ is carried by the harmonic pieces
$h_\mu$ of the gauge fields $A_\mu(x)$. These $h_\mu$ are of the same type
as the local \CS-like terms encountered previously in four dimensions
\cite{K00,K98}. Indeed,
the \CS~one-form for an one-dimensional Abelian $U(1)$ gauge field $a(x)$ is
given by
\begin{equation}\label{eq:omegaCS}
\omega_{\rm \,CS} [a] \equiv (2\pi)^{-1} \,   a(x) \,\rmd x  \;.
\end{equation}
One possible two-dimensional \CS-like term is then the average over the
$x^2$ coordinate
of $2\pi i$ times the genuine \CS~term for the $x^1$ space $S^1$, namely
\beqa\label{eq:CSlike}
 && \Gamma_{\rm \,CS-like,1}^{\,S^1 \times S^1}[A\,] \equiv
    \int_{0}^{L} \frac{\rmd x^2}{L} \left( \,2\pi i\int_{S^1}
              \omega_{\rm \,CS}[A_1]\right) \nonumber \\
 &&   = i\int_{0}^{L} \rmd x^1 \int_{0}^{L} \rmd x^2\,
              A_1(x^1,x^2)/L = 2\pi i h_1 \;,
\eeqa
where $h_1$ is defined by Eq. (\ref{eq:Adecomposed}). The other two-dimensional
\CS-like term (based on the genuine \CS~term for the $x^2$ space)
equals $2\pi i h_2$. Hence, \CS-like terms play a role for the CPT anomaly
in both two and four dimensions. There is, however, a difference,
in that the four-dimensional \CS-like term immediately affects the  gauge
field propagation, with the vacuum becoming optically active \cite{K00,CFJ90}.

In closing, we remark that the CPT non\-in\-var\-ian\-ce
found here appears to be not directly related to the purely gravitational
anomaly which afflicts Weyl fermions in two dimensions
($4\,k+2$ dimensions in general) \cite{AW84}.
The gravitational anomaly (breakdown of general coordinate invariance)
of the two-dimensional 345-model, say,
shows up for deviations from the Euclidean
metric $g_{\mu\nu}$ $=$ $\delta_{\mu\nu}$,
but in our case the metric is perfectly Euclidean and, still,
the effective gauge field action $\Gamma^{\rm PP}[A]$ is CPT noninvariant.
Instead of local spacetime
fluctuations, it is the spacetime topology (and spin structure)
that is relevant to the CPT anomaly.
The CPT anomaly resembles in this respect the
so-called topological Casimir effect \cite{noteCasimir}.

\section*{Acknowledgments}
This work was started during the sabbatical leave of
one of us (F.R.K.) and the hospitality of the
High-Energy Theory group at the Niels Bohr Institute is gratefully
acknowledged.
J.N. is supported by the Japan Society for the Promotion of
Science. 


\begin{references}
\bibitem{noteDPP} To arrive at Eq. (\ref{eq:DPP})
         we have chosen factors $P[b_1]$ $=$ $+\frac{1}{2}$ and
         $P[b_2]$ $=$ $+\frac{1}{2}$ in Eq. (2.18) of Ref. \cite{IN99} but other
         signs are also possible, as long as they are fixed once and for all.
\bibitem{notePhase}  For the 345-model with PP spin structure, the
         \cdet~from Ref. \cite{KNN97} would not be invariant under arbitrary
         large gauge transformations. This shows that the extra phase factor
         $\exp \left( i\pi(h _ 1 - h _2)/2 \right)$ in Eq. (\ref{eq:DPP}),
         which traces back to the local counterterm of Ref. \cite{IN99}, also plays a
         role in maintaining the Abelian gauge invariance of the full
         \cdet~(\ref{eq:D345}).
\bibitem{noteTheta} The relevant properties of $\vartheta (z;\tau)$
         are its periodicity under $z$ $\rightarrow$ $z+1$ and
         quasi-periodicity under $z$ $\rightarrow$ $z+\tau$, together with the
         symmetry $\vartheta (-z;\tau)$ $=$ $\vartheta (z;\tau)$.
         We take the opportunity to correct Eq. (4.13) of Ref. \cite{IN99},
         which must have an extra factor $\exp\left(i\pi n_1 n_2 \right)$ on the \rhs.
\bibitem{noteEigenval} For constant gauge potentials
                   $A_\mu(x)$ $= $ $2\pi h_\mu/L$, the single chiral determinant
                   $D^{\rm PP}[h_1,h_2]$  from Eq. (\ref{eq:pathintegral}) is,
                   formally, proportional to
                   $\prod \left(n_1+h_1+i n_2 +i h_2\right)$, with the
                   product running over the integers $n_1$ and $n_2$.
                   The CPT-odd factor $(h_1+i h_2)$ is manifest for
                   $n_1$ $=$ $n_2$ $=$ $0$
\pagebreak 
                   (which corresponds to the
                   eigenvector $\psi(x)$ $=$ ${\rm const.}$ of the Weyl operator),
                   whereas the other contributions combine into CPT-even
                   factors whose product still needs to be regularized appropriately.
                   The full \cdet~(\ref{eq:D345}) of the 345-model then has factors
                   $3\,(h_1+i h_2)$, $4\,(h_1+i h_2)$, and $5\,(h_1-i h_2)$,
                   which combine to give the result (\ref{eq:D345linear}).
\bibitem{notePV} For the two-dimensional chiral $U(1)$ gauge theory on a torus, we
 have also calculated the \cdet~for constant gauge potentials
 $A_1(x)$ $= $ $2\pi h_1/L$ and $A_2(x)$ $=$ $0$,
 using the  partial regularization method adopted in Ref. \cite{K00}.
 For this setup, the Fourier modes of the $x^2$ direction decouple and
 each of them can be regarded as
 a one-dimensional Dirac fermion in $x^1$ space.
 The regularization  method then amounts to introducing
 one-dimensional Pauli-Villars fields over $x^1$ space
 for each Fourier mode from the $x^2$ direction.
 The result for the \cdet~differs in general from the one obtained in
 Ref. \cite{IN99} and behaves differently under modular transformations.
 In particular, CPT violation is observed
 if the \bcs~on the fermion fields are taken to be periodic in the $x^2$ direction
 and either periodic or antiperiodic in the $x^1$ direction,
 whereas the result from Ref. \cite{IN99} is CPT violating only for the doubly-periodic
 (PP) spin structure,  as shown by Eq. (\ref{eq:345CPT}).
 The apparent regularization dependence of the theory deserves further study; cf.
 Ref \cite{JR85}.  Still, the existence of a CPT anomaly for the PP spin structure is
 essentially independent of the
 ultraviolet regularization, as discussed below Eq. (\ref{eq:D345linear})
 and in \cite{noteEigenval}.
\end{references}
\end{document}